\documentclass[12pt,tightenlines,preprintnumbers,%
superscriptaddress,nofootinbib]{revtex4}
\newcommand{\PRE}[1]{{#1}} 

\usepackage{bm} \usepackage{epsfig}

\newcommand{\eqref}[1]{Eq.~(\ref{#1})}

\begin{document}

\preprint{MIT/CTP-3701}  

\title{

The Universe is a Strange Place\footnote{Public lecture given at Lepton-Photon 2005, Uppsala, Sweden July 2005.  Earlier I used the same title for a quite different talk, astro-ph/0401347. }

}

\PRE{\vspace*{.5in}}

\author{Frank Wilczek%
\PRE{\vspace*{.2in}}
} 
\affiliation{Center for Theoretical Physics, Department of Physics,
Massachusetts Institute of Technology, Cambridge, Massachusetts 02139,
USA
\PRE{\vspace*{.5in}}
}


\begin{abstract}

Our understanding of ordinary matter is remarkably accurate and complete, but it is based on principles that are very strange and unfamiliar.   As I'll explain, we've come to understand matter to be a Music of the Void, in a remarkably literal sense.   Just as we physicists finalized that wonderful understanding, towards the end of the twentieth century, astronomers gave us back our humility, by informing us that ordinary matter -- what we, and chemists and biologists, and astronomers themselves, have been studying all these centuries constitutes only about 5\% of the mass of the universe as a whole.   I'll describe some of our promising attempts to rise to this challenge by improving, rather than merely complicating, our description of the world.

\end{abstract}

\maketitle

\PRE{\newpage}

In a lecture of one hour I will not be able to do justice to all the ways in which the Universe is a strange place.  But I'll share a few highlights with you.  

\section{Interior Strangeness}

First I'd like to talk about how strange I am.   Oh, and you too, of course.   That is, I'd like to describe how strange and far removed from everyday experience is the accurate picture of ordinary matter, the stuff that we're made of, that comes from modern physics.   

I think that 10,000 years from now, our descendants will look back on the twentieth century as a very special time. It was the time when humankind first came to understood how matter works, indeed what it {\it is}.   

By 1900, physicists had a great deal of Newtonian-level knowledge.    The paradigm for this level of understanding is Newton's account of the motion of planets, moons, and comets in the solar system.  Newtonian-level knowledge takes the form: if bodies are arranged with given positions and velocities at time $t_0$, then the laws of physics tell you what their positions and velocities will be at any other time.  Newton's theory is extremely accurate and fruitful.  It led, for example, to the discovery of a new planet  -- Neptune -- whose gravity was necessary to account for anomalies in motion of the known planets.   (We're facing similar problems today, as I'll describe later.  Neptune was the ``dark matter'' of its day.)  

But nothing in Newton's theory prescribes how many planets there are.  Nothing predicts their masses, or their distances from the Sun.   For planetary systems all that freedom is a good thing, we know now, because astronomers are discovering other systems of planets around other stars, and these solar systems are quite different from ours, but they still obey Newton's laws.   

But no such freedom is observed for the building blocks of ordinary matter.   Those building blocks come in only a few kinds, that can only fit together in very restricted ways.   Otherwise there could not be such a subject as chemistry, because each sample of matter would be different from every other sample.   Pre-modern physics could not account for that fact, and therefore it could not even begin to account for the specific chemical, electrical, and mechanical properties of matter.

The turning point came in 1913, with Bohr's model of the hydrogen atom.   Bohr's model brought quantum ideas into the description of matter.   It pictured the hydrogen atom as analogous to a simplified solar system (just one planet!) held together by electric rather than gravitational forces, with the proton playing the role of the sun and the electron the role of the planet.  The crucial new idea in Bohr's model  is that only certain orbits are allowed to the electron: namely, oribts for which the angular momentum is a whole-number multiple of Planck's constant.   When the atom exchanges energy with the outside world, by emitting or absorbing light, the electron jumps from one orbit to another, decreasing its energy by emitting a photon or increasing its energy by absorbing one.   The frequency of the photon, which tells us what color of light it conveys, is proportional to its energy according to the Planck-Einstein relation $E=h\nu$, where $E$ is the energy, $\nu$ is the frequency, and $h$ is Planck's constant.  Because the allowed orbits form a discrete set, rather than a continuum, we find discrete spectral lines corresponding to the allowed energy changes.   

Bohr's model predicts the colors of the spectral lines remarkably accurately, though not perfectly. When Einstein learned of Bohr's model, he called it ``the highest form of musicality in the sphere of thought".   I suppose Einstein was alluding here to the ancient idea of ``Music of the Spheres'', which has enchanted mathematically inclined mystics from Pythagoras to Kepler.   According to that idea the planets produce a kind of music through their stately periodic motions,  as the strings of a violin do through their periodic vibrations.   Whether the ``Music of the Spheres'' was intended to be actual sounds or something less tangible, a harmonious state induced in the appreciative mind, I'm not sure.  But in Bohr's model the connection is close to being tangible: the electron really does signal its motion to us in a sensory form, as light, and the frequencies in the line spectrum are the tonal palette of the atomic instrument.   

So that's one way in which Bohr's model is musical.  Another is in its power to suggest meanings far beyond its actual content.   Einstein sensed immediately that Bohr's introduction of discreteness into the description of matter, with its constraint on the possible motions, intimated that the profound limitations of pre-modern physics which I just described would be overcome -- even though, in itself, Bohr's model only supplied an approximate description of the simplest sort of atom.  

In one respect, however, Einstein was wrong.  Bohr's model is definitely {\it not\/} the highest form of musicality in the sphere of thought.    The theory that replaced Bohr's model, modern quantum mechanics, outdoes it by far.   

In modern quantum mechanics, an electron is no longer described as a particle in orbit.  Rather, it is described by a vibrating wave pattern in all space. The equation that describes the electron's wave function, Schr\"odinger's equation, is very similar to the equations that describe the vibrations of musical instruments.  In  Schr\"odinger's account light is emitted or absorbed when the electron's vibrations set the electromagnetic field -- aether, if you like -- in motion, by the same sort of sympathetic vibration that leads to the emission of sound by musical instruments, when their vibrations set air in motion.   These regular, continuous processes replace the  mysterious ``quantum jumps'' from one orbit to another that were assumed, but not explained, in Bohr's model.   

A major reason that physicists were able to make rapid progress in atomic physics, once Schr\"odinger found his equation, is that they were able to borrow techniques that had already been used to analyze problems in sound production and music.   Ironically, despite his well-known love of music, Einstein himself never accepted modern quantum mechanics. 

After the consolidation of atomic physics in the early 1930s, the inner boundary of physics shrank by a factor of a hundred thousand.   The challenge was to understand the tiny atomic nuclei, wherein are concentrated almost all the mass of matter.   The nuclei could only held together by some new force, which came to be called the strong force, since gravity is much too feeble and electrical forces are both too feeble and of the wrong sign to do the job (being repulsive, not attractive).  Experimenters found that it was useful to think of nuclei as being built up from protons and neutrons, and so the program initially was to study the forces between protons and neutrons.   You would do this by shooting protons and neutrons at each other, and studying how they swerved.   But when the experiments were done, what emerged was not just existing particles swerving around in informative directions, but a whole new world containing many many new particles, a Greek and Roman alphabet soup containing among others $\pi, \rho, K, \omega, \phi$ mesons and $\Delta, \Lambda, \Sigma, \Omega$ baryons and their antiparticles, that are unstable, but otherwise bear a strong family resemblance to protons and neutrons.   So the notion of using protons and neutrons as elementary building blocks, bound together by forces you would just go ahead and measure, became untenable.  

I'll skip over the complicated story of how physicists struggled to recover from this confusing shock, and just proceed to the answer, as it began to emerge in the 1970s, and was firmly established during the 1990s. The building blocks of nuclei are quarks and gluons. Add electrons and photons, and you have all the building blocks you need for atoms, molecules, ions, and indeed all ordinary matter.

Quarks once had a bad reputation, because for many years attempts to produce them in the laboratory, or to find them anywhere, failed.  People went so far as to search with great care through cosmic ray tracks, and even in moon rocks.    And gluons had no reputation at all.    But now we know that quarks and gluons are very real.  You can see them, quite literally --- once you know what to look for!  Here
\begin{figure}[h]
\begin{center}
\includegraphics[scale=.30]{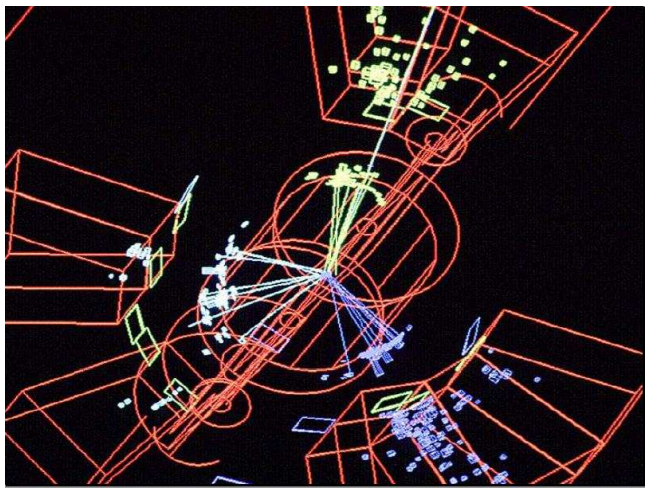}
\caption{A three-jet event at LEP.  Courtesy L3 collaboration.}
\label{3jets}
\end{center}
\end{figure}
(Fig.~\ref{3jets}) is a picture of a quark, an antiquark, and a gluon.  What is shown here is the result of colliding electrons and positrons at very high energy.    It was taken at the Large Electron Positron collider (LEP) at CERN, near Geneva, which operated through the 1990s.   You see that the particles are emerging in three groups, which we call jets.  I want to convince you that one of these jets represents a quark, one an antiquark, and one a gluon.  

The key idea for justifying this interpretation is {\it asymptotic freedom}, for which the Nobel Prize was awarded in 2004.   Asymptotic freedom says that an energetic quark (or antiquark or gluon) will frequently emit soft radiation, which does not significantly change the overall flow of energy and momentum; but only rarely emit hard radiation, which does produce changes in the flow.  Here's what asymptotic freedom means, concretely, in the LEP situation.   Right after the electron and positron annihilate, their energy is deposited into a quark-antiquark pair moving rapidly in opposite directions.    Usually the quark and antiquark will only emit soft radiation.     In that case, which occurs about 90\% of the time, we will have two jets of particles, moving in opposite directions, whose total energy and momentum reflect those of the original quark and antiquark.   More rarely, about 10\% of the time, there will be a hard radiation, where a gluon is emitted.   That gluon will then initiate a third jet.   And 10\% of the 10\%, that is 1\%, of the time a second hard radiation will occur, and we'll have four jets, and so forth.   Having invested hundreds of millions of Euros to construct the LEP machine, physicists exploited it intensely, producing many millions of collisions, working up to about one bang per buck.   So it was possible to study the properties of these multi-jet events, to see how likely it was for them to emerge at different angles and to share the available energy in various ways, and thereby to test precisely whether they behave in the way that our basic theory of quarks and gluons, namely quantum chromodynamics (QCD), predicts.   It works.  That's why I can tell you, with complete confidence, that what you're seeing in the picture is a quark, an antiquark, and a gluon.   We see them not as particles in the conventional sense, but through the flows they imprint on the visible particles.   

There's another aspect of this business that I think is extremely profound, though it is so deeply ingrained among physicists that they tend to take it for granted.   It is this: These observations provide a wonderfully direct illustration of the probabilistic nature of quantum theory.   At LEP experimentalists just did one thing over and over again, that is collide electrons and positrons.  We know from many experiments that electrons and positrons have no significant internal structure, so there's no question that when we make these collisions we really are doing the same thing over and over again.    If the world were governed by deterministic equations, then the final result of every collision would be the same, and the hundreds of millions of Euros would have had a very meagre payoff.  But according to the principles of quantum theory many outcomes are possible, and our task is to compute the probabilities.   That task, computing the probabilities, is exactly what QCD accomplishes so successfully.   

By the way, one consequence of the probabilistic nature of our predictions is that while I can tell you that you're seeing a quark, an antiquark and a gluon, I can't say for sure which is which!

So we know by very direct observations that quarks and gluons are fundamental constituents of matter.   QCD proposes that gluons and quarks are all we need to make protons, neutrons, and all the other strongly interacting particles.    That's an amazing claim, because there's a big disconnect between the properties of quarks and gluons and the properties of the things they are supposed to make.    

Most notably, gluons have strictly zero mass, and the relevant quarks have practically zero mass, but together, it's claimed, they make protons and neutrons, which provide overwhelmingly most of the mass of ordinary matter.    That claim flies in the face of the ``conservation of mass'' principle that Newton used as the basis of classical mechanics and Lavoisier used as the foundation of quantitative chemistry.   Indeed, before 1905, this idea of getting mass from no mass would have been inconceivable.    But then Einstein discovered his second law.

My friend and mentor Sam Treiman liked to relate his experience of how, during World War II, the U.S. Army responded to the challenge of training a large number of radio engineers starting with very different levels of preparation, ranging down to near zero.  They designed a crash course for it, which Sam took, and a training manual, which Sam loved, and showed me .  In that training manual, the first chapter was devoted to Ohm's three laws.  
The Army designed Ohm's first law is $V=IR$.  Ohm's second law is $I=V/R$.  I'll leave it to you to reconstruct Ohm's third law.

Similarly, as a companion to Einstein's famous equation $E=mc^2$ we have his second law, $m=E/c^2$.  

All this isn't quite as silly as it may seem, because different forms of the same equation can suggest very different things.  The great theoretical physicist Paul Dirac described his method for making discoveries as ``playing with equations''.   

The usual way of writing the equation, $E=mc^2$, suggests the possibility of obtaining large amounts of energy by converting small amounts of mass.  It brings to mind the possibilities of nuclear reactors, or bombs.   
In the alternative form
  $m=E/c^2$, Einstein's law suggests the possibility of explaining mass in terms of energy.  That is a good thing to do, because in modern physics energy is a more basic concept than mass.   It is energy that is strictly conserved, energy that appears in the laws of thermodynamics, energy that appears in Schr\"odinger's equation.   Mass, by contrast, is a rather special, technical concept -- it labels irreducible representations of the Poincar\'{e} group (I won't elaborate on that.)

Actually, Einstein's original paper does not contain the equation $E=mc^2$, but rather $m=E/c^2$.  So maybe I should have called $m=E/c^2$ the zeroth law, but I thought that might be confusing.  The title of the original paper is a question: ``Does the Inertia of a Body Depend Upon its Energy Content?'' From the beginning, Einstein was thinking about the foundations of physics, not about making bombs.   I think he would have been delighted to learn that our answer to his question is a resounding ``Yes!''   Not only does the inertia of bodies depend on its energy content; for ordinary matter most of inertia {\it is\/} the energy associated with moving quarks and gluons, themselves essentially massless, following $m=E/c^2$.   Who knows, it might even have inspired him to accept modern quantum mechanics.  

To solve the equations of QCD, and identify the different ways in which quarks and gluons can organize themselves into the particles we observe, physicists have pushed the envelope of modern computing.   They sculpt upwards of $10^{30}$ protons and neutrons into a massively parallel computer, which runs at teraflop speeds -- that is, a thousand billion, or $10^{12}$ multiplications of big numbers {\it per second\/} for months, that is $10^7$ seconds.   All to calculate what every single proton does in $10^{-24}$ seconds,  that is figure out how to arrange its quarks and gluons efficiently, to get the minimum energy.   

\begin{figure}[h]
\begin{center}
\includegraphics[scale=.4]{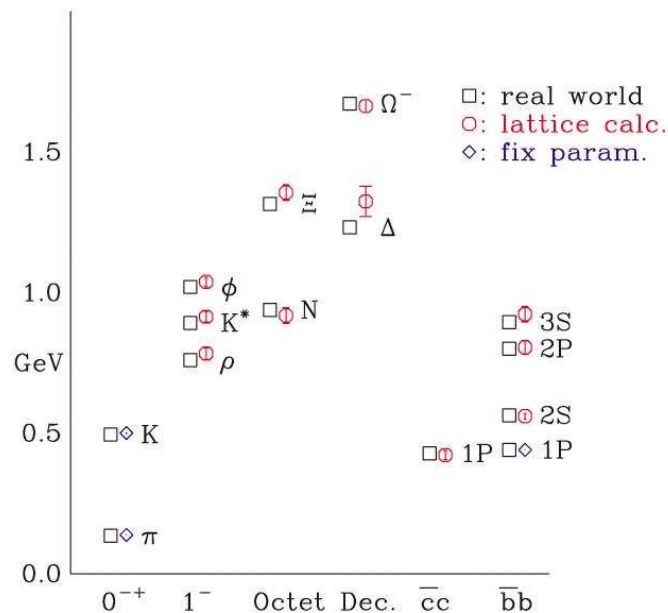}
\caption{Hadron masses as computed numerically in QCD.   Courtesy D. Toussaint.}
\label{masses}
\end{center}
\end{figure}

Evidently there's room for improvement in our methods of calculation.   But already the results are most remarkable.   They are displayed in Fig.~\ref{masses}.
I think what you see in that modest-looking plot is one of the greatest scientific achievements of all time.   We start with a very specific and mathematically tight theory, QCD.   An objective sign of how tight the theory is, is that just a very few parameters have to be taken from experiment.  Then everything else is fixed, and must be computed by algorithms the theory supplies, with no room for maneuver or fudge factors.   Here three parameters were fixed, indicated by the diamonds, by matching the masses of the $\pi$ and $K$ mesons and a splitting among heavy quark mesons (I won't enter the technicalities) and then all the other calculated masses are displayed as circles, with line intervals indicating the remaining uncertainty in calculation (due to computer limitations).   As you see, they agree quite well with the measured values, indicated as squares.   

What makes this {\it tour de force\/} not only impressive, but also historic, is that one of the entries, $N$, means ``nucleon'', that is, proton or neutron.  So QCD really does account for the mass of protons and neutrons, and therefore of ordinary matter, starting from ideally simple elementary objects, the quarks and gluons, which themselves have essentially zero mass.   In this way, QCD fulfills the promise of Einstein's second law.  

Another important aspect of Figure \ref{masses} is what you {\it don't\/} see.   The computations do not produce particles that have the properties of individual quarks or gluons.   Those objects, while they are the building block, are calculated never to occur as distinct individual particles.   They are always confined within more complex particles -- or, as we've seen, reconstructed from jets.    This confinement property, which historically made quarks and gluons difficult to conceive and even more difficult to accept, is now a calculated consequence of our equations for their behavior.  

Thus our theory corresponds to reality, in considerable detail, wherever we can check it.   Therefore we can use it with some confidence to explore domains of reality that are extremely interesting, but difficult to access directly by experiment.   

\begin{figure}[h]
\begin{center}
\includegraphics[scale=.17]{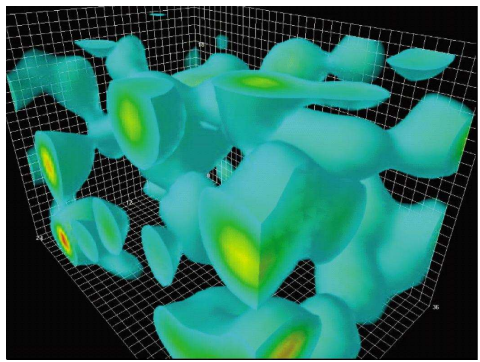}
\caption{Gluon fields fluctuating in Void.  Courtesy D. Leinweber.}
\label{gluonLavaLamp}
\end{center}
\end{figure}

I think that for the future of physics, and certainly for the future of this lecture, the most profound and surprising result to emerge from late twentieth-century physics may be the realization that what we perceive as empty space is in reality a highly structured and vibrant dynamical medium.   Our eyes were not evolved to see that structure, but we can use our theories to calculate what empty space might look like if we had eyes that could resolve down to distances of order $10^{-14}$ centimeters, and times of order $10^{-24}$ seconds.   Derek Leinweber in particular has put a lot of effort into producing visualizations of the behavior of quark and gluon fields, and I highly recommend his website www.physics.adelaide.edu.au/theory/ staff/leinweber/VisualQCD/QCDvacuum/\\ welcome.html as a source of enlightenment.  Figure~\ref{gluonLavaLamp} shows gluon fields as they fluctuate in ``empty'' space.  I want to emphasize that this is not a free fantasy, but part of the calculation that leads to Fig.~\ref{masses}  (for experts: what is shown is a smoothed distribution of topological charge).

The different particles we observe are the vibration patterns that are set up in the medium of ``empty'' space, let's call it Void, when it is disturbed in different ways.  Stable particles such as protons correspond to stable vibration patterns; unstable particles correspond to vibration patterns that hold together for a while, then break apart.   This is not a metaphor, it is our most profound understanding.   Rather it is more familiar and conventional ideas about matter that are metaphors, for this deeper reality.

\begin{figure}[h]
\begin{center}
\includegraphics[scale=.40]{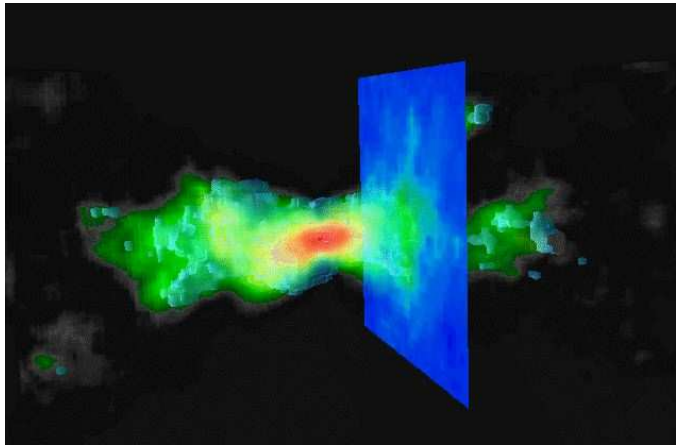}
\caption{Disturbance in Void from injecting quark-antiquark pair.  Courtesy G. Kilcup.]}
\label{pion}
\end{center}
\end{figure}

Indeed, the way Figure \ref{masses} was produced, and more generally the way we compute the properties of matter from first principles, is to introduce some disturbance in Void, let it settle down for a while, and observe what kind of stable or long-lived patterns emerge.   An example is shown in Fig.~\ref{pion}. 
Here a quark and an antiquark are inserted on the left, the medium responds, and a stable vibration pattern emerges (the plane shows a slice of time).  This picture is obtained by averaging over the fluctuations that would occur in the absence of the quark and antiquark, so the stuff in Figure \ref{gluonLavaLamp} has been subtracted; we're interested in the net disturbance.   This is how we produce a $\pi$ meson for numerical study.    The picture for a proton would look similar; you'd get it by studying the disturbance when you introduce three quarks, instead of a quark and antiquark.   

We go from these vibration patterns to masses by combining Einstein's second law $m= E/c^2$ with the Einstein-Planck relation $E= h \nu$ between the energy of a state and the frequency at which its wave function vibrates.   Thus
$$
m ~=~ E/c^2 ~=~ h\nu/c^2
$$
or alternatively
$$
\nu ~=~ mc^2/h.
$$
Thus masses of particles correspond in a very direct and literal way to the frequencies at which the Void vibrates, when it is disturbed.   That is how we calculate them.  The ancient ``Music of the Spheres'' was an inspiring concept, but it never corresponded to ideas that are very precise or impressive.  Now we have a Music of the Void, which I trust you'll agree is all three of these things.

\section{Our Strange Surroundings}

Just as we physicists were finally consolidating this powerful understanding of ordinary matter, astronomers made some amazing new discoveries, to help us maintain our humility.   They discovered that the sort of matter we've been familiar with, made from electrons, photons, quarks, and gluons --- the stuff we're made of, the stuff of chemistry and biology, the stuff of stars and nebulae --- makes up only 5\% of the Universe by mass.   The remainder consists of at least new substances.  There's 25\% in something we call dark matter, and 70\% in what we call dark energy.  

Very little is known about dark matter, and even less about dark energy.  One thing we do know is that neither of them is really dark.   They're transparent.   They neither emit nor absorb light to any significant extent --- if they did, we'd have discovered them a long time ago.   In fact, dark matter and dark energy seem to interact very feebly not only with photons, but with ordinary matter altogether.    They've only been detected through their gravitational influence on the (ordinary) kind of matter we do observe.   

Other things we know: Dark matter forms clumps, but not such dense clumps as ordinary matter.   Around every visible galaxy that's been carefully studied we find an extended halo of dark matter, whose density falls off much more slowly than that of ordinary matter as you recede from the center.   It's because it is more diffusely distributed that averaged over the Universe as a whole the dark matter has more total mass than ordinary matter, even though ordinary matter tends to dominate in the regions where it is found. 

Dark energy doesn't seem to clump at all.  It is equally dense everywhere, as far as we can tell, as if it is an intrinsic property of space-time itself.   Most strange of all, dark energy exerts negative pressure, causing the expansion of the Universe to accelerate.   

With that, I've basically told you everything we know about dark matter and dark energy.  It's not very satisfying.   We'd like to know, for example, if the dark matter is made of particles.  If so, what do those particles weigh?  Do they really not interact with matter at all, except by gravitation, or just a little more feebly than we've been sensitive to so far?   Do dark matter particles interact strongly with each other?  Can they collide with one another and annihilate?   

How do you find answers to questions like those?   One way, of course, is to do experiments.   But the experiments have not borne fruit so far, as I mentioned, and as a  general rule it's difficult to find something if you don't know what you're looking for.    There's another way to proceed, which has a great history in physics.  That is, you can improve the equations of fundamental physics.   You can try to make them more consistent, or to improve their mathematical beauty.  

For example, in the middle of the nineteenth century James Clerk Maxwell studied the equations of electricity and magnetism as they were then known, and discovered that they contained a mathematical inconsistency. At the same time Michael Faraday, a self-taught genius of experimental physics who did not have great skill in mathematics, had developed a picture of electric and magnetic phenomena that suggested to Maxwell how he might fix the inconsistency, by adding another term to the equations.   When Maxwell added this term, he found that the new equations had solutions where changing electric fields induce magnetic fields, and vice versa, so that you could have self-supporting waves of electromagnetic disturbance traveling through space at the speed of light. Maxwell proposed that his electromagnetic disturbances in fact {\it are\/} light, and in this way produced one of the great unifications in the history of physics. As a bonus, he predicted that there could be electromagnetic waves of different wavelength and frequency, which would in effect be new forms of ``light'', not visible to human eyes.  Waves of this sort were finally produced and detected by Heinrich Hertz in 1888; today of course we call them radio waves, and also microwaves, gamma rays, and so on. Another example came around 1930, when Paul Dirac worked to improve Erwin Schr\"odinger's equation for the quantum mechanical wave function of electrons.   Schr\"odinger's equation, as I mentioned before, made a big logical improvement on Bohr's model and is quite successful in giving a first account of atomic spectra. We still use it today. But Schr\"odinger's equation has a severe theoretical flaw: it is not consistent with special relativity. In 1928 Dirac invented an improved equation for electrons, that implements quantum dynamics and is also consistent with special relativity.  Some of the solutions of Dirac's equation correspond closely to solutions of solutions of Schr\"odinger's equation, with small corrections.  But Dirac's equation has additional solutions, that are completely new.  At first it was quite unclear what these solutions meant physically, but after some struggles and false starts in 1932 Dirac put forward a convincing interpretation.  The new solutions describe a new kind of matter, previously unsuspected.   Dirac predicted the existence of antielectrons, or positrons. Within a few months, the experimentalist Carl Anderson found positrons, by studying cosmic rays.   It was the first example of antimatter.   Nowadays positrons are used for medical purposes (Positron Emission Tomography, or PET), and many other kinds of antimatter have been discovered.   

Today we have several good new ideas for how to improve the equations of physics.  I'll mention a few momentarily.   But first let me make a preliminary observation:  Because we understand so much about how matter behaves in extreme conditions, and because that behavior is remarkably simple, we can work out the cosmological consequences of changes in our fundamental equations.  If our suggestion for improving the equations predicts new kinds of particles, we can predict the abundance with which those particles would be produced during the big bang.   If any of the particles are stable, we will predict their present cosmological density.  If we're lucky, we might find that one of our new particles is produced in the right amount, and has the right properties, to supply the astronomers' dark matter.   
 
\begin{figure}[h]
\begin{center}
\includegraphics[scale=.30]{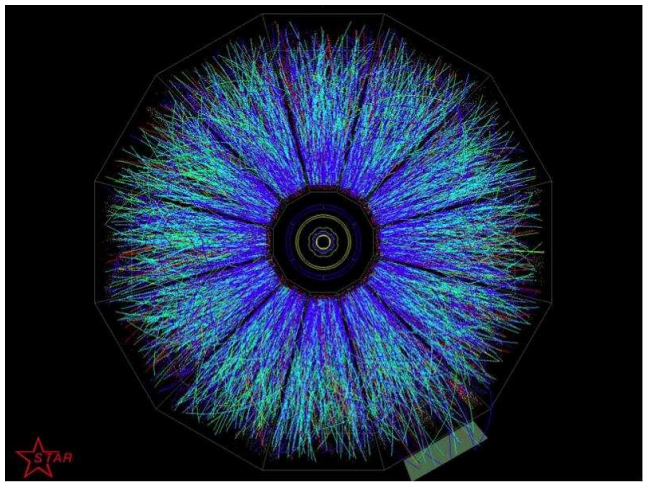}
\caption{Particles emerging from a violent collision of gold nuclei, which reproduces physical conditions close to the big bang.  Courtesy STAR collaboration.}
\label{fireball}
\end{center}
\end{figure}

Recently experimenters have been testing our theoretical understanding of the early universe in a most remarkable way. By colliding gold nuclei at extremely high energies they create, for very brief times and in a very small volume (around $10^{-20}$ seconds, and $10^{-12}$ centimeters), conditions of temperature and density in terrestrial laboratories similar to those that last occurred throughout the universe a hundredth of a second or so after the Big Bang. The ashes of these tiny fireballs are thousands of particles, as shown in Fig.~\ref{fireball}. 
 It looks very complicated, and in many ways it is, but our theories makes many predictions about the overall flow of energy and the properties of the most energetic particles that can be compared with observations, and those predictions work pretty well, so we're encouraged, and emboldened.

So, how do we go about improving our equations?  
Over the course of the twentieth century, symmetry has been immensely fruitful as a source of insight into Nature's basic operating principles.   QCD, in particular, is constructed as the unique embodiment of a huge symmetry group, local $SU(3)$ color gauge symmetry (working together with special relativity, in the context of quantum field theory).   As we try to discover new laws that improve on what we know, it seems good strategy to continue to use symmetry as our guide.    This strategy has led physicists to several compelling suggestions.  
Let me very briefly mention four of them:
\begin{enumerate}
\item We can combine our theories of the strong, weak, and electromagnetic interactions into a single unified theory, by extending the symmetries that form the basis of these theories into a larger symmetry, that contains all of them (and more).   This is known as grand unification.  Grand unification predicts the existence of new particles and new phenomena, including mass for neutrinos (which has been observed) and instability of protons (not yet observed).   
\item We can extend the space-time symmetry of special relativity to include mixing of space and time with additional quantum dimensions.  This is known as supersymmetry.  Supersymmetry predicts the existence of a whole new world of particles.  Each currently known particle will have a heavier superpartner, with different spin.  
\item We can enhance the equations of QCD by adding a symmetry that explains why the strong interactions exhibit no preferred arrow of time.   This leads us to predict, by quite subtle and advanced arguments, the existence of a new kind of extremely light, extremely feebly interacting particle, the axion. 
\item We can enhance the symmetry of our equations for the weak and electromagnetic interactions, and 
achieve a partial unification, by postulating the existence of a universal background field, the so-called Higgs condensate, that fills all space and time.   There is already a lot of indirect evidence for that idea, but we'd like to produce the new particles, the Higgs particles, that this field is supposed to be made of.     
\end{enumerate}
Each of these items leads into a beautiful story and an active area of research, and it's somewhat of a torture for me to restrain myself from saying a lot more about them, but  time forbids.  I'll just describe one particular line of ideas, that ties together the first two of these items with the dark matter problem, and that will soon be tested decisively.

Both QCD and the modern theory of electromagnetic and weak interactions are founded on similar mathematical ideas.  The combination of theories gives a wonderfully economical and powerful account of an astonishing range of phenomena.   It constitutes what we call the Standard Model.   Just because it is so concrete and so successful, the Standard Model can and should be closely scrutinized for its aesthetic flaws and possibilities.  In fact, the structure of the Standard Model gives powerful suggestions for
its further fruitful development. They are a bit technical to describe, but I'll say a few words, which you can take as poetry if not as information.  

The product structure
$SU(3)\times SU(2) \times U(1)$ or the gauge symmetry of the Standard Model, the reducibility of the fermion representation
(that is, the fact that the symmetry does not make connections linking
all the fermions), and the peculiar values of the hypercharge quantum numbers assigned to the known particles all suggest the desirability of a larger symmetry.   
The devil is in the details, and it is not at all automatic that the
superficially complex and messy observed pattern of matter will fit neatly into a simple
mathematical structure. But, to a remarkable extent, it does.

There seems to be a big problem with implementing more perfect symmetry among the different interactions, however.   The different interactions, as observed, do not have the same overall strength, as would be required by the extended symmetry.   The strong interaction, mediated by gluons, really is observed to be much stronger than the electromagnetic interaction, mediated by photons.   That makes it difficult, on the face of it, to interpret gluons and photons as different aspects of a common reality.   

But now we should recall that empty space is a dynamical medium, aboil with quantum fluctuations.   Gluons or photons see particles not in their pristine form, but rather through the distorting effects of this unavoidable medium.   Could it be that when we correct for the distortion, the underlying equality of different interactions is revealed?   

\begin{figure}[h]
\begin{center}
\includegraphics[scale=.20]{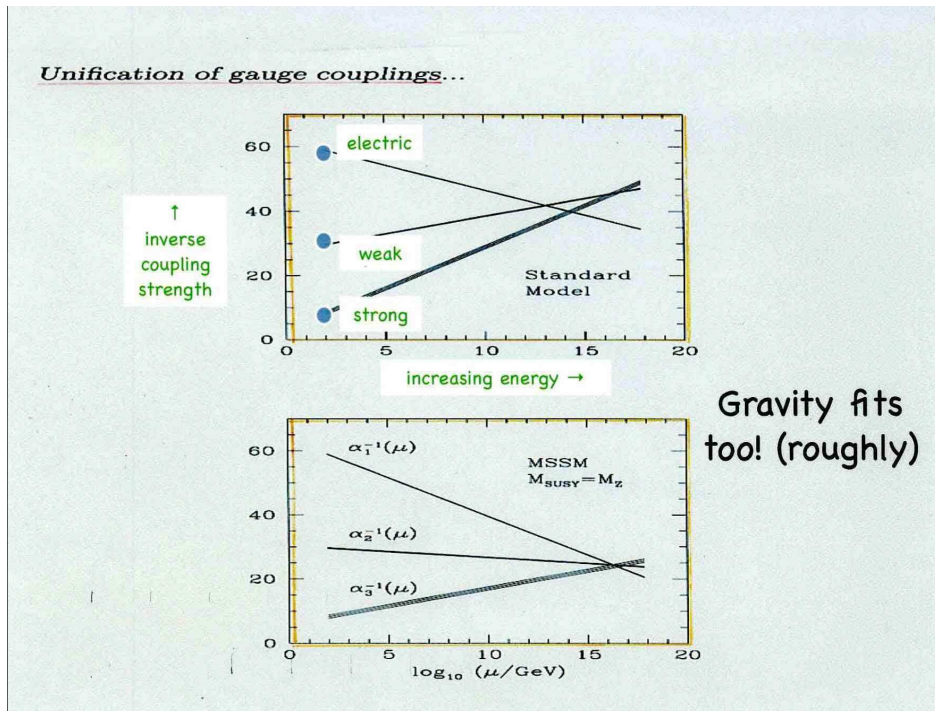}
\caption{Unification of couplings using the currently known particles (upper panel) and with low-energy supersymmetry (lower panel).}
\label{running}
\end{center}
\end{figure}

To try out that idea, we have to extend our theory to distances far smaller than, or equivalently energies far larger than, we have so far accessed experimentally.   Fig.~\ref{running} gives a sense of what's involved. 
The left-most part of the graph, with the discs, represents our actual measurements. It's a logarithmic scale, so each tick on the horizontal axis means a factor of ten in energy. Building an accelerator capable of supplying one more factor of ten in energy will cost a few billion Euros. After that it gets {\it really\/} difficult. So the prospects for getting this information directly are not bright. Nevertheless, it's interesting to calculate, and if we do that we find some very intriguing results. 

In correcting for the medium, we must include fluctuations due to all kinds of fields, including those that create and destroy particles we haven't yet discovered. So we have to make some hypothesis, about what kind of additional particles there might be. The simplest hypothesis is just that there are none, beyond those we know already. Assuming this hypothesis, we arrive at the calculation displayed in the top panel of Fig.~\ref{running}. You see that the approach of coupling strengths to a unified value is suggested, but it is not quite accurately realized. 

Physicists react to this near-success in different ways.   One school says that near-success is still failure, and you should just give up this dream of unification, that the world is bound to be a lot more complicated.   Another school  says that there is some truth in the basic idea, but the straightforward extrapolation of physics as we know it (based on quantum field theory) to such extreme energies and distances is wrong.   You might have to include the effect of extra dimensions, or of strings, for example.    So you should be grateful that this calculation works as well as it does, and wait for  revolutionary developments in physics to teach you how to improve it.

Either or both of these reactions might turn out to be right. But I've long advocated a more definite and (relatively) conservative proposal that still seems promising, and I'd like to mention it now, having warned you that not all my colleagues have signed on to it, by any means.  
It is based on yet another way to improve the equations of physics, known as low-energy supersymmetry.

As the name suggests, supersymmetry involves expanding the symmetry of the basic equations of physics. This proposed expansion of symmetry goes in a different direction from the enlargement of gauge symmetry, which we've just been considering. Supersymmetry connects particles having the same color charges and different spins, whereas expanded gauge symmetry changes the color charges while leaving spin untouched. Supersymmetry expands the space-time symmetry of special relativity.   

In order to implement low-energy supersymmetry, we must postulate the existence of a whole new world of heavy particles, none of which has yet been observed directly. There is, however, a most intriguing indirect hint that this idea may be on the right track. If we include the particles needed for low-energy supersymmetry, in their virtual form, into the calculation of how couplings evolve with energy, then accurate unification is achieved! This is shown in the bottom panel of Fig.~\ref{running}.

Among the many new particles, one is particularly interesting. It is electrically neutral and has no color charge, so it interacts very weakly with ordinary matter. It lives a very long time -- longer than the current lifetime of the Universe. Finally, if we calculate how much of it would be produced during the big bang, we find that it supplies roughly the right density to supply the dark matter. All this adds up to suggest that maybe this new particle is supplying the dark matter astronomers have discovered.    

By ascending a tower of speculation, involving now both extended gauge symmetry and extended space-time symmetry, we seem to break though the clouds, into clarity and breathtaking vision. Is it an illusion, or reality? That question creates a most exciting situation for the Large Hadron Collider (LHC), due to begin operating at CERN in 2007. For that great accelerator will achieve the energies necessary to access the new world of heavy particles, if it exists.  How the story will play out, only time will tell.  In any case, I think it is fair to say that the pursuit of unified field theories, which in past (and many present) incarnations has been vague and barren of testable consequences, has in the circle of ideas I've been describing here attained entirely new levels of concreteness and fertility.

 \section{Three Great Lessons}

Now I'm done with what I planned to tell you today about the strangeness of the Universe. I think it's appropriate to conclude by connecting these grand considerations about the nature of reality to human life. So I'll conclude by drawing what I feel are three great lessons --- I'm not sure whether I should call them moral, philosophical, or spiritual lessons --- from the scientific results I've described.
 
\begin{description}
\item[1.] {\bf The part of the world we understand is strange and beautiful.}  We, and all the things we deal with in everyday life, are Music of the Void.
\item[2.] {\bf If we work to understand, we can understand.}   Using hands and minds evolved for quite other purposes, by diligent labor and honest thought, we have come to comprehend vastly alien realms of the infinite and the infinitesimal.  
\item[3.] {\bf We still have a lot to learn.}
\end{description}
 
\bigskip
\bigskip

\*{Acknowledgment}

\bigskip

The work of FW is supported in part by funds provided by
the U.S. Department of Energy under cooperative research agreement
DE-FC02-94ER40818.

\end{document}